\documentclass[aip, apl, reprint, superscriptaddress,showpacs,amsmath,amssymb, twocolumn, floatfix]{revtex4-1}

\usepackage{graphicx}
\usepackage{dcolumn}
\usepackage{bm}
\usepackage{color}
\usepackage{ulem}
\usepackage[utf8]{inputenc}

\definecolor{KommentarPhilip}{gray}{.8}  

\newcommand{\nphot}{\bar{n}_{\textrm{c}}} 
\newcommand{\omegac}{\omega_{\textrm{c}}} 

\newcommand{\gtc}{g_{\textrm{q}}} 
\newcommand{\gmV}{g_{\textrm{m0}}}

\newcommand{\omegap}{\omega_{\textrm{p}}} 

\newcommand{\omegaqb}{\omega_{\textrm{s}}} 

\newcommand{\omegad}{\omega_{\textrm{d}}} 
\newcommand{\Papp}{P_{\textrm{appl}}}      

\newcommand{\deltatc}{\Delta_{\textrm{qc}}} 

\newcommand{\omegat}{\omega_{\textrm{q}}} 

\newcommand{\Omegam}{\Omega_{\textrm{m}}} 
\newcommand{\Gammam}{\Gamma_{\textrm{m}}} 
\newcommand{\GammaEff}{\Gamma_{\textrm{eff}}} 

\newcommand{\xemia}{x_{\textrm{EMIA}}}

\newcommand{\xqb}{x_{\textrm{qb}}}

\begin{document}


\title{Ultrawide-range photon number calibration using a hybrid system combining nano-electromechanics and superconducting circuit quantum electrodynamics}

\author{Philip Schmidt}
\email{Philip.Schmidt@wmi.badw.de}
\affiliation{Walther-Mei{\ss}ner-Institut, Bayerische Akademie der Wissenschaften, Walther-Mei{\ss}ner-Str. 8, 85748 Garching, Germany}%
\affiliation{Physik-Departement, Technische Universit{\"a}t M{\"unchen}, James-Franck-Str. 1, 85748 Garching, Germany}
\affiliation{Nanosystems Initiative Munich, Schellingstra{\ss}e 4, 80799 M{\"unchen}, Germany}%

\author{Daniel Schwienbacher}%
\affiliation{Walther-Mei{\ss}ner-Institut, Bayerische Akademie der Wissenschaften, Walther-Mei{\ss}ner-Str. 8, 85748 Garching, Germany}%
\affiliation{Physik-Departement, Technische Universit{\"a}t M{\"unchen}, James-Franck-Str. 1, 85748 Garching, Germany}
\affiliation{Nanosystems Initiative Munich, Schellingstra{\ss}e 4, 80799 M{\"unchen}, Germany}%

\author{Matthias Pernpeintner}%
\affiliation{Walther-Mei{\ss}ner-Institut, Bayerische Akademie der Wissenschaften, Walther-Mei{\ss}ner-Str. 8, 85748 Garching, Germany}%
\affiliation{Physik-Departement, Technische Universit{\"a}t M{\"unchen}, James-Franck-Str. 1, 85748 Garching, Germany}
\affiliation{Nanosystems Initiative Munich, Schellingstra{\ss}e 4, 80799 M{\"unchen}, Germany}%

\author{Friedrich Wulschner}
\affiliation{Walther-Mei{\ss}ner-Institut, Bayerische Akademie der Wissenschaften, Walther-Mei{\ss}ner-Str. 8, 85748 Garching, Germany}%
\affiliation{Physik-Departement, Technische Universit{\"a}t M{\"unchen}, James-Franck-Str. 1, 85748 Garching, Germany}

\author{Frank Deppe}%
\affiliation{Walther-Mei{\ss}ner-Institut, Bayerische Akademie der Wissenschaften, Walther-Mei{\ss}ner-Str. 8, 85748 Garching, Germany}%
\affiliation{Physik-Departement, Technische Universit{\"a}t M{\"unchen}, James-Franck-Str. 1, 85748 Garching, Germany}

\author{Achim Marx}%
\affiliation{Walther-Mei{\ss}ner-Institut, Bayerische Akademie der Wissenschaften, Walther-Mei{\ss}ner-Str. 8, 85748 Garching, Germany}%

\author{Rudolf Gross}%
\affiliation{Walther-Mei{\ss}ner-Institut, Bayerische Akademie der Wissenschaften, Walther-Mei{\ss}ner-Str. 8, 85748 Garching, Germany}%
\affiliation{Physik-Departement, Technische Universit{\"a}t M{\"unchen}, James-Franck-Str. 1, 85748 Garching, Germany}
\affiliation{Nanosystems Initiative Munich, Schellingstra{\ss}e 4, 80799 M{\"unchen}, Germany}%

\author{Hans Huebl}%
\email{huebl@wmi.badw.de}
\affiliation{Walther-Mei{\ss}ner-Institut, Bayerische Akademie der Wissenschaften, Walther-Mei{\ss}ner-Str. 8, 85748 Garching, Germany}%
\affiliation{Physik-Departement, Technische Universit{\"a}t M{\"unchen}, James-Franck-Str. 1, 85748 Garching, Germany}
\affiliation{Nanosystems Initiative Munich, Schellingstra{\ss}e 4, 80799 M{\"unchen}, Germany}%

\date{\today}

\begin{abstract}
We present a hybrid system consisting of a superconducting coplanar waveguide resonator coupled to a nanomechanical string and a transmon qubit acting as nonlinear circuit element. We perform spectroscopy for both the transmon qubit and the nanomechanical string. Measuring the ac-Stark shift on the transmon qubit as well as the electromechanically induced absorption on the string allows us to determine the average photon number in the microwave resonator in both the low and high power regimes. In this way, we measure photon numbers that are up to nine orders of magnitude apart. We find a quantitative agreement between the calibration of photon numbers in the microwave resonator using the two methods. Our experiments demonstrate the successful combination of superconducting circuit quantum electrodynamics and nano-electromechanics on a single chip.
\end{abstract}
\maketitle
The field of optomechanics allows to investigate the interaction of light with mechanical degrees of freedom. It enables the optical readout of the mechanical displacement \cite{Braginsky1995} as well as the control of the mechanical state. Over the past decade, optomechanics has been successfully used to study the interplay between mechanical modes and quantized electromagnetic waves in a resonator on the quantum level \cite{Aspelmeyer2016, Wollman2015, Pirkkalainen2015, Lecocq2015, Lei2016, Vivoli2016, Hofer2016, Abdi2016b, Borkje2016, Abdi2015}. 
One successful experimental implementation is based on superconducting circuits as they, straightforwardly, can be operated in the resolved sideband limit \cite{Regal2008}.

In addition, superconducting nano-electromechanical circuits are compatible with the field of circuit quantum electrodynamics (cQED) \cite{Schoelkopf2008} regarding fabrication technology, operation temperature and frequency range. In cQED, the strong \cite{Wallraff2004, Blais2004} and ultra-strong coupling regimes \cite{Niemczyk2010, Baust2016, FornDiaz2016, Yoshihara2017} have been achieved and the generation of non-classical states of microwave light is well established \cite{Johansson2006, Schuster2007, Hofheinz2008, Eichler2011, Menzel2012, Chen2014}. Therefore, the combination of  nano-electromechanics with cQED is an ideal approach to delve into the quantum nature of mechanical motion.

Recent experiments \cite{Pirkkalainen2015c, OConnell2009} show that the combination of a superconducting qubit, a microwave resonator and a nanomechanical element can enhance the phonon-photon interaction, allow for the controlled preparation of non-classical phonon states, and enable entanglement generation. 
One envisaged state preparation protocol is the generation of a non-classical microwave state in a microwave resonator coupled to both a qubit and a mechanical resonator. It makes use of a well-defined qubit state and its transfer to the mechanical system via a red-sideband drive pulse. For this, one critical parameter is the average photon number in the microwave resonator.

Here, we present an experimental study of a hybrid quantum system consisting of a transmon qubit, a doubly clamped high-Q nanomechanical string resonator, and a superconducting microwave resonator. We show that the average photon number determined from the ac-Stark shift and electromechanically induced absorption (EMIA) measurements are in good quantitative agreement.

The average photon number inside a $\lambda/2$ microwave resonator with symmetric input/output (total) coupling rate $\kappa_{\textrm{ext}}$ is given by \cite{Aspelmeyer2014, Clerk2010}
\begin{equation}
\nphot =  \frac{2 \Papp}{\hbar\omega_{\textrm{p}}\left(\kappa^2 + 4\Delta_{\textrm{p}}^2\right)}\underbrace{\Lambda\kappa_{\textrm{ext}}}_{\equiv x}.
\label{eq:nc2}
\end{equation}
Here,  $\hbar$ is the reduced Planck constant, $\Delta_{\textrm{p}} \equiv \omegap - \omegac$ denotes the detuning between the probe tone frequency $\omegap$ and the resonant frequency $\omegac$. Additionally, $\kappa \equiv \kappa_{\textrm{int}} + \kappa_{\textrm{ext}}$ describes the total loss rate of the microwave resonator given by the sum of the internal $(\kappa_{\textrm{int}})$ and external $(\kappa_{\textrm{ext}})$ losses. The applied power $\Papp$ describes the total calibrated output power of the microwave sources before sending it to the dilution refrigerator. Since the attenuation $\Lambda$ of the microwave lines and $\kappa_{\textrm{ext}}$ can only be estimated, we introduce the product $\Lambda \cdot \kappa_{\textrm{ext}}$ as a calibration factor $x$ in Eq.$\,(\ref{eq:nc2})$. We demonstrate that $x$ can be quantified and corroborate its value via two independent approaches: (i) We measure the ac-Stark shift of the qubit transition frequency as a function of $\Papp$ in the dispersive regime \cite{Walls2008, Blais2004} which we will call $\xqb$ and (ii) we measure EMIA resulting from the  electromechanical interference effect between the anti-Stokes field of the coupled electromechanical resonator system and a probe field \cite{Teufel2011-2, Hocke2012, Zhou2013, Bagci2014, Singh2014} determining $\xemia$.
\begin{figure*}
  \includegraphics[scale=1]{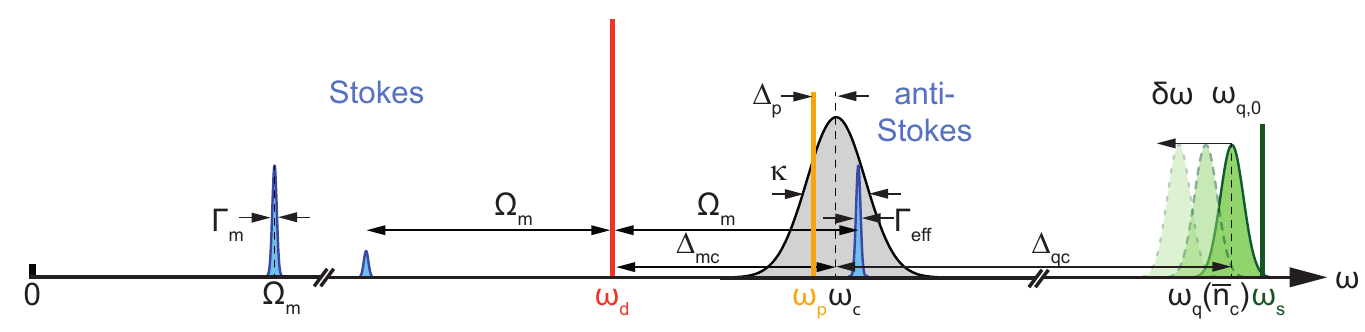}
        \caption{\textit{Eigenfrequencies and excitation tones of the hybrid system consisting of a superconducting microwave resonator, a transmon qubit, and a nanomechanical string resonator.} The mechanical element (blue) has an eigenfrequency $\Omegam$. For the so-called red-sideband configuration, the frequency $\omegad$ of the microwave drive tone is chosen to position the anti-Stokes field at or close to the resonance frequency of the microwave resonator $\omegac$. When the frequency $\omegap$ of the probe tone is resonant with the anti-Stokes field, electromechanically induced absorption is observed. The average photon number $\nphot$ causes an ac-Stark shift $\delta\omega (\nphot)$ of the bare qubit transition frequency $\omega_{q,0}$. The photon number dependent qubit frequency, $\omegat(\nphot) = \omega_{\textrm{q,}0}-\delta\omega$, is determined using a spectroscopy tone with frequency $\omega_{\textrm{s}}$.}
        \label{EmitScheme} 
\end{figure*}

Taking the linewidth $\kappa$ of the microwave resonator into account, the ac-Stark shift for a transmon qubit is given by \cite{Koch2007, Jan351, Jan353}
\begin{equation}
\delta \omega =  2 \frac{\gtc^2} {\deltatc}\frac{\alpha}{\alpha+\deltatc}\nphot(\kappa),
\label{eq:acstark}
\end{equation}
when probing the microwave resonator on resonance $(\Delta_{\textrm{p}}=0)$. Here, the coupling between the transmon and the microwave resonator is $\gtc$ and $\deltatc  =\omegat-\omegac$ (cf. Fig.\,\ref{EmitScheme}). The non-linearity of the transmon qubit is defined by $\alpha$, quantifying the deviation in energy of the second mode from twice the ground mode. 
Evidently, we can determine $\nphot$ and, in turn, the calibration factor $\xqb$ by measuring $\delta \omega$ if we know the qubit parameters. 

Next, we turn to EMIA which is known to result in an increase of the linewidth $\Gammam$ of the mechanical oscillator, leading to an effective linewidth $\GammaEff$ given by \cite{Weis2010, Zhou2013,Singh2014}
\begin{equation}
\GammaEff =  \Gammam\left (1 + \frac{4 \gmV^2}{\kappa \Gammam} \frac{2 \Papp\xemia}{\hbar\omegad\left(\kappa^2 + 4\Delta_{\textrm{mc}}^2\right)} \right ).
\label{eq:EMIAFit}
\end{equation}
Thus, measuring $\GammaEff$ as a function of $\Papp$ allows us to determine the calibration factor $\xemia$, if we know the relevant parameters of the mechanical oscillator and the electromechanical vacuum coupling constant $\gmV$. For this experiment, we chose $\omegad=\omegac -\Omegam$, i.e., $\Delta_{\textrm{mc}}= -\Omegam$ (cf. Fig.$\,\ref{EmitScheme}$).
\begin{figure}
  \includegraphics[scale=1]{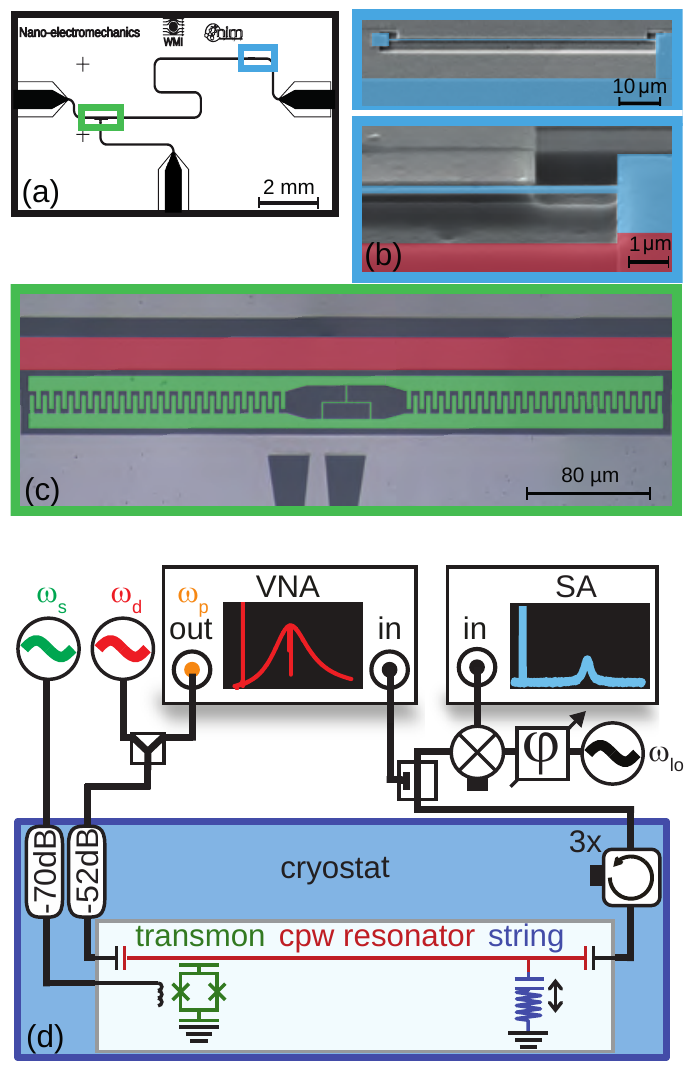}
        \caption{\textit{Sample layout, images, and spectroscopy setup.} Panel (a) shows the chip layout including the $\lambda/2$ coplanar waveguide microwave resonator, the transmon qubit (green box), and the doubly clamped aluminum nanostring resonator (blue box). Panel (b) shows a scanning electron micrograph of the $60\,\mu$m long nanostring including an enlarged view of the region close to the right clamp. The enlarged image is tilted to view the successful release. An additional antenna structure is placed close to the qubit (green) in panel (c). In panel (d) multiple microwave sources act as red-sideband drive tone $\omegad$, qubit spectroscopy tone $\omega_{\textrm{s}}$, and local oscillator $\omega_{\textrm{lo}}$. The vector network analyzer (VNA) supplies the probe tone and allows direct analysis of the transmitted signal. In addition, a spectrum analyzer is used to analyze the sideband fluctuations of the mechanical element. The output signal of the microwave resonator is preamplified with a cryogenic HEMT amplifier at 4\,K, followed by post-amplification at room temperature (not shown).}
        \label{fig:Setup}
\end{figure}

To quantitatively compare the resonator photon numbers determined with the two methods described above, we fabricate a hybrid device consisting of a superconducting microwave resonator, a transmon qubit and a doubly clamped nanomechanical string resonator, as depicted in Fig.\,\ref{fig:Setup}. All these parts consist of superconducting aluminum thin films deposited by electron beam evaporation on a single crystalline silicon substrate. Patterning of the microwave and nanomechanical resonator is achieved by electron beam lithography and a lift-off process. After Al deposition, the sample is annealed at $300\,^\circ$C for 30 minutes to generate a high tensile stress in the aluminum thin film. Then, the transmon qubit is defined again by electron beam lithography and fabricated using a two-angle shadow evaporation (see Ref.~\onlinecite{Jan351} and Ref.~\onlinecite{Jan353}). In the last step, we release the nanostring resonator by reactive ion etching and critical point drying.

The $60\,\mu$m long, $120\,$nm thick, and $230\,$nm wide nano-string resonator has a mass of about $2\,$pg. It is separated by a $160\,$nm gap from the ground plane resulting in  $\gmV/2\pi=0.31\,$Hz. At the experimental temperature of $T_{\textrm{cryo}}\simeq50\,$mK we observe a mechanical resonance frequency of $\Omegam/2\pi = 3.15018\,$MHz, corresponding to a zero point fluctuation amplitude of $35\,$fm. The low intrinsic mechanical linewidth of $\Gammam/2\pi = 12.4\,$Hz corresponds to a thermal coherence time of $38\,\mu$s.

Tuning the transmon qubit to its minimum frequency, far away from the resonator frequency, we find the bare microwave resonator frequency $\omegac/2\pi = 5.875\,$GHz. Its linewidth depends on the setpoint of the transmon qubit and the photon occupation inside, see supplementary material for details.

As further detailed in the supplementary material, we find for the transmon qubit an eigenfrequency of $\omegat /2\pi = 7.916\,$GHz at the sweet spot, corresponding to a detuning of $\deltatc / 2\pi = 2.056\,$GHz, a transmon nonlinearity of $\alpha / 2\pi = -188\,$MHz, and a transmon-resonator coupling of $\gtc / 2\pi = 134\,$MHz.
\begin{figure}
  \includegraphics[scale=1]{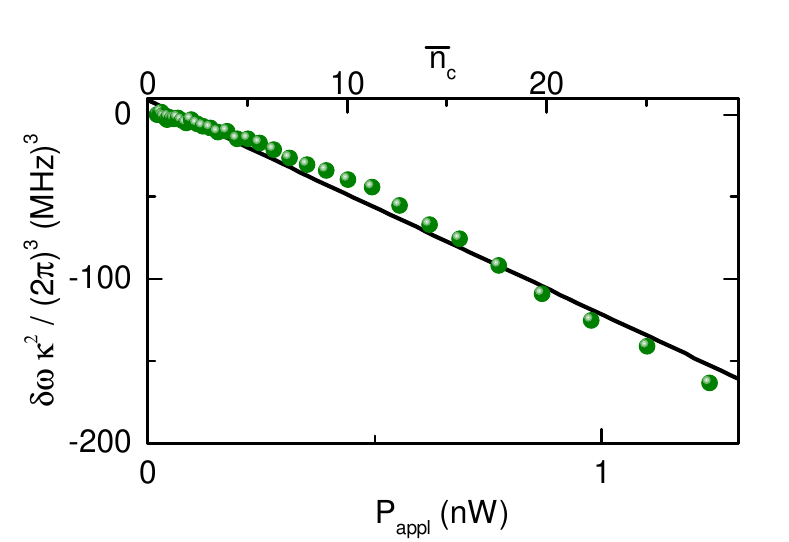}
        \caption{\textit{Ac-Stark shift of the transmon.} The product $\delta\omega \kappa^2$ (green dots) is plotted as a function of the applied microwave power $\Papp$. A linear model [see Eq.\,(\ref{eq:acstark})] is fitted to the data (black solid line). The determined photon numbers are shown on the top axis.}
        \label{fig:ACStark}
\end{figure}

For the measurement of the qubit transition frequency as a function of $\Papp$ in the few photon regime, we tune the qubit to its sweet spot by an applied magnetic field. We then perform two-tone spectroscopy by driving the transmon qubit via its antenna while probing the resonator transmission with a microwave tone of varying power $\Papp$. In this way, we obtain the qubit frequency as a function of $\Papp$. In addition, we determine the microwave resonator linewidth as $\nphot$ is influenced by this parameter. The product $\delta\omega\cdot\kappa^2$ is shown in Fig.$\,\ref{fig:ACStark}$. As expected [c.f. Eq.$\,(\ref{eq:acstark})$], this product linearly depends on $\Papp$. We find a slope of $-(2\pi)^3\cdot (1.30\pm0.03)\cdot10^{20}\,/$s$^{3}$nW. Combining this slope, as well as Eq.$\,(\ref{eq:nc2})$, and the system parameters, we obtain $\xqb = (5.65\pm0.23)\,$s$^{-1}$. Note that we used probe powers corresponding to $\nphot \le 28$, well below the critical photon number of $\bar{n}_{\textrm{crit}}=\deltatc^2 / (2\gtc)^2 \approx 60$, set by the assumptions of the dispersive limit \cite{Bois2009, Raftery2014}.

\begin{figure}
  \includegraphics[scale=1]{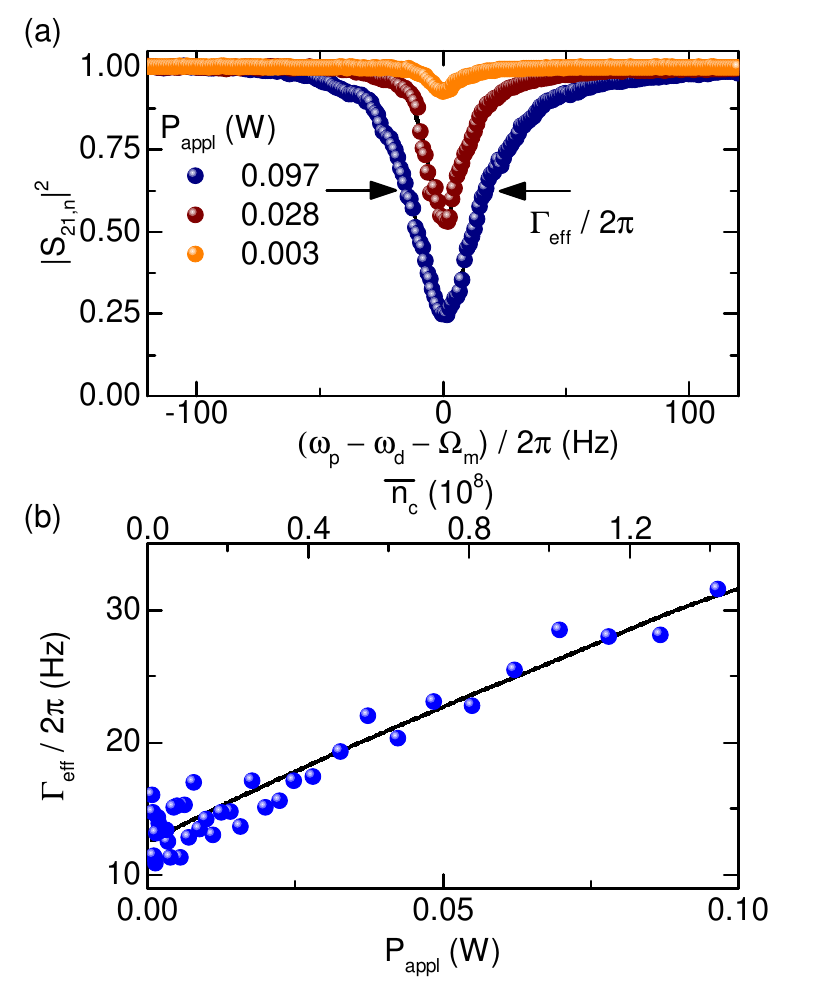}
        \caption{\textit{EMIA spectroscopy on the electromechanical hybrid system.} Panel (a) shows the normalized EMIA signature (dotted) for three drive powers $\Papp$. The EMIA dip deepens and widens with increasing drive power. Lorentzian models are fitted to the data (black solid lines) determining $\GammaEff$. Panel (b) displays the extracted effective linewidth $\GammaEff$ (blue dots) as a function of the red-sideband drive amplitude at a fridge temperature of about $50\,$mK. Using the model described in the main text (black solid line), we find the calibration factor $\xemia$ and the photon numbers in the microwave resonator (top axis).}
        \label{TwoToneResults}
\end{figure}

For the determination of the resonator photon numbers at higher occupations, we turn to the two-tone EMIA spectroscopy scheme (cf. Fig.$\,\ref{EmitScheme}$). We set the red-sideband drive tone to $\omegad=\omegac-\Omegam$ and probe the anti-Stokes field  with the probe tone $\omegap$, close to $\omegac$. Depending on the red-sideband drive amplitude, i.e. $\nphot$, we obtain the spectra depicted in Fig.\,\ref{TwoToneResults}(a). This figure shows the EMIA, which manifests as an additional absorption around $\omegap = \omegad+\Omegam$. By fitting a Lorentzian lineshape to the EMIA data, we extract the effective interference linewidth $\GammaEff$.
A quantitative analysis of the transmission data shows an EMIA signature with a minimum corresponding to $25\,\%$ of the unperturbed microwave transmission parameter $|S_{\textrm{21,n}}|^2$.

Figure\,\ref{TwoToneResults}$\,$(b) displays the extracted EMIA linewidth $\GammaEff$ as a function of the  applied red-sideband drive power, confirming the linear increase predicted by Eq.$\,(\ref{eq:EMIAFit})$. The calibration factor $\xemia$ is determined by fitting Eq.$\,(\ref{eq:EMIAFit})$ to the data. Having characterized the sample parameters, the calibration factor $\xemia$ and the intrinsic linewidth $\Gammam$ are the remaining free parameters in the model. We  obtain $\xemia = (5.41\pm0.25)\,$s$^{-1}$ and $\Gammam/2\pi = (12.4\pm0.3)\,$Hz. Using these results in combination with Eq.$\,(\ref{eq:nc2})$, we can determine the average photon numbers in the microwave resonator in a range from about $10^{6}$ to $10^{8}$ photons.

In conclusion, we have successfully implemented a superconducting coplanar microwave resonator coupled to a transmon qubit and a nanomechanical string. Both the coupled transmon-MW resonator system and the nano-electromechanical system were investigated using microwave spectroscopy. These experiments allowed us to calibrate the MW resonator photon numbers by measuring the ac-Stark shift of the transmon qubit in the range from $0.7$ to $28$ photons and showed a calibration factor of $\xqb = (5.65\pm0.23)\,$s$^{-1}$. For higher photon numbers, we used EMIA spectroscopy of the nanomechanical string to probe the resonator in a population range from $1.4\times 10^{6}$ to $1.4\times 10^{8}$ photons. By analyzing the linewidth of the transmission signature, we found a calibration factor $\xemia=(5.41\pm0.25)\,$s$^{-1}$. Both attenuation coefficients $\xqb$ and $\xemia$ agree within $5\%$. Please also note that these two methods determine average resonator photon numbers that are up to nine orders of magnitude apart.

The implementation of a transmon qubit, a high-quality nanostring resonator, and a microwave resonator on a single chip represents an important step towards the realization of quantum information storage in the vibrational degree of freedom of a mechanical element. Doubly clamped string resonators are interesting candidates in this context, due to their high mechanical quality factors, above $10^5\,$, corresponding to a thermal coherence time from micro- to milliseconds at a moderate dilution fridge temperature of $50\,$mK. 

\begin{acknowledgments}
This project has received funding from the European Union's Horizon 2020 research and innovation program under grant agreement No 736943. We gratefully acknowledge valuable scientific discussions with J. Goetz, S. Weichselbaumer, and E. Xie.
\end{acknowledgments}

\appendix

\section{Transmon qubit}
\label{sec:TransmonCharacterization}
\begin{figure}
\includegraphics[scale=1]{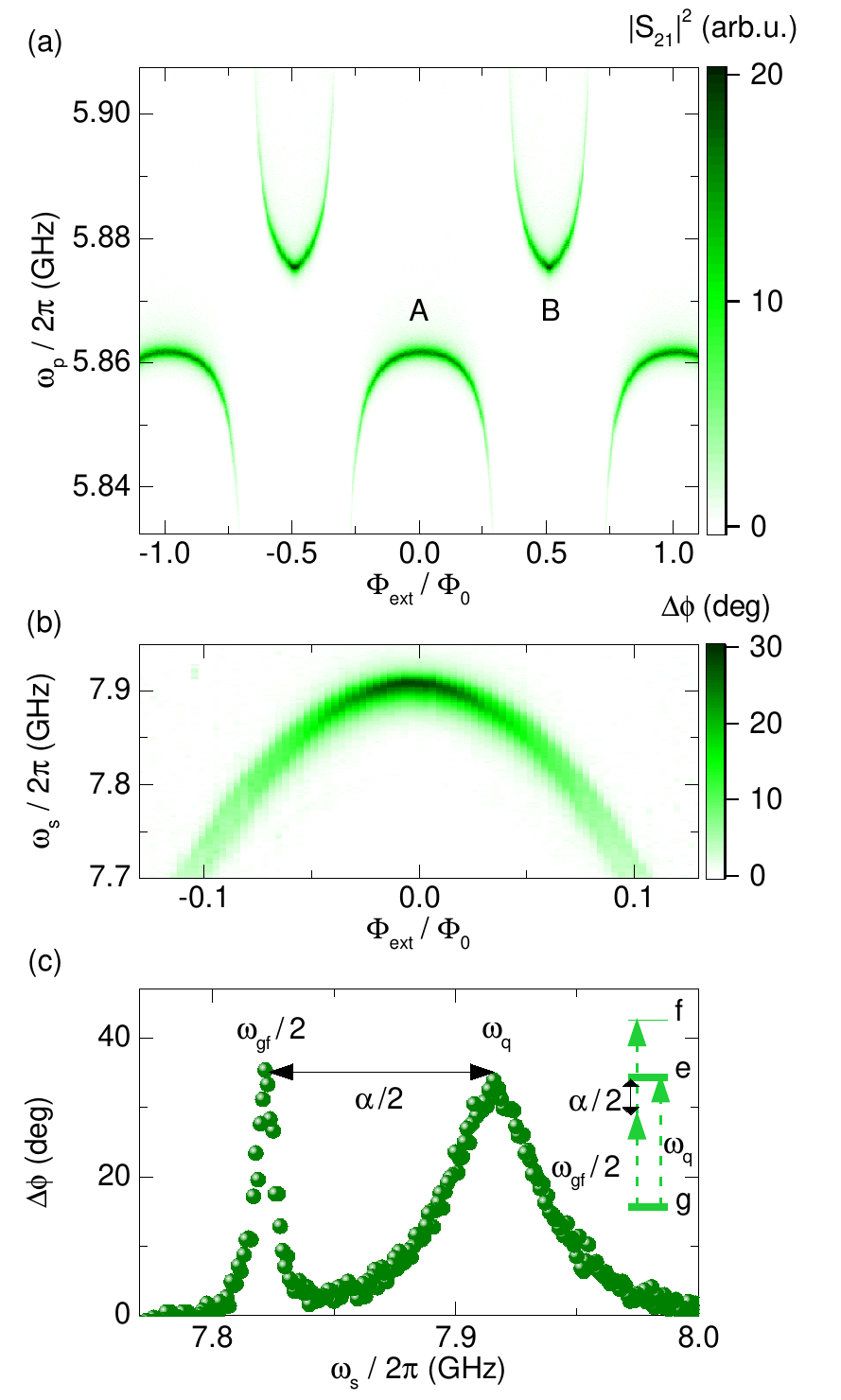}
\caption{\textit{Spectroscopy of the coupled resonator-transmon qubit system} (a) Microwave transmission $|S_{21}|^2$ as a function of the applied magnetic flux $\Phi_{\textrm{ext}}$. A periodic behavior is found for integer flux ratios. When the transmon qubit is in resonance with the microwave resonator an avoided crossing is observed. (b) Transmon  transition frequency $\omegat$ as a function of the applied flux, measured by two-tone spectroscopy (phase change of the probe signal, right axis). The maximum frequency is observed at zero field. (c) High power spectroscopy data of the transmon qubit at the sweet spot $\Phi_{\textrm{ext}}/\Phi_0 = 0$ (green dots) show the single photon $|g\rangle\leftrightarrow|e\rangle$ transition at $\omegat$ as well as the two photon  $|g\rangle\leftrightarrow|f\rangle$ transition at $\omega_{\textrm{gf}}$, as schematically depicted on the right.}
\label{fig:CavityResponse}
\end{figure}
The transmon qubit is positioned at the electric field anti-node of the coplanar waveguide resonator and is capacitively coupled to it. The qubit transition frequency can be varied by applying a magnetic flux $\Phi$ to the dc-SQUID forming the tunable Josephson junction of the transmon qubit. 

Figure\,\ref{fig:CavityResponse}(a) shows the transmission of a weak probe tone $\omegap$ through the microwave resonator as a function of the applied magnetic flux for an average photon number of 1.7. At $\Phi_{\textrm{ext}} /\Phi_0 = \pm 0.28$, we find a pronounced anti-crossing demonstrating the strong coupling between the transmon qubit and the resonator. From the peak separation at the avoided crossing, for occupations below one photon on average, we find a coupling strength of $\gtc / 2\pi = (134.1\pm2.3)\,$MHz.
 Additionally, we find the expected periodic flux dependence of the qubit transition frequency. When the transmon qubit is tuned to its minimum frequency, e.g. at $\Phi_{\textrm{ext}}/\Phi_0=0.5$, the detuning between the qubit and the resonator is so large that the uncoupled resonator frequency can be determined to $\omegac/2\pi = 5.875\,$GHz, with a linewidth of $\kappa / 2\pi = (1.468\pm0.022)\,$MHz.

When the qubit is far detuned from the resonator and therefore in the dispersive regime, the probed resonance frequency of the microwave resonator depends on the qubit state \cite{Walls2008, Blais2004}. Thus, driving the qubit with $\omegaqb$, allows to perform a two-tone spectroscopy of the qubit, as shown in Fig.\,\ref{fig:CavityResponse}(b) and (c).
From panel (b) we determine a qubit frequency ($|g\rangle\leftrightarrow|e\rangle$ transition) of $\omegat/2\pi=7.916$\,GHz at the sweet spot of $\Phi_{\textrm{ext}}/\Phi_0=0$. 

To access the transmon anharmonicity $\alpha$ we increase the amplitude of the drive tone at $\Phi_{\textrm{ext}} / \Phi_0=0$. For high drive powers two- and multi-photon processes become visible \cite{Bishop2008, Braumueller2015}. In particular, we observe the two-photon transition $|g\rangle\leftrightarrow|f\rangle$ at $\omega_{\textrm{gf}}/4\pi=(7.8220\pm0.0005)\,$GHz corresponding to a transmon qubit anharmonicity of $\alpha / \hbar = 2\omegaqb - \omega_{\textrm{gf}}= -2\pi\cdot(188\pm1)\cdot10^6\,$s$^{-1}$. For transmon qubits the negative anharmonicity is equivalent to the charging energy $(-\alpha = E_{\textrm{C}})\,$ \cite{Koch2007}.
Via $\omegac = \sqrt{8E_{\textrm{C}}E_{\textrm{J}}} / \hbar\,$ \cite{Koch2007} we find an $E_{\textrm{J}}/E_{\textrm{C}}$ ratio of 222 for the transmon qubit.

\section{Calibration of electromechanical coupling strengths}
Next, we turn to the characterization of the nano-electromechanical system. We use the thermal fluctuations of the nanostring, similar to Refs.\,[\onlinecite{Zhou2013, Hocke2012, Gorodetsky2010, Weber2016}], to determine the electromechanical vacuum coupling constant $\gmV/2\pi=(0.308\pm0.004)\,$Hz. In detail, we use a frequency-modulated drive tone set to $\omegad = \omegac$ to probe the frequency fluctuations of the microwave resonator, caused by the thermal motion of the nanostring resonator. The transmitted signal is down-converted using a homodyne setup and analyzed with a spectrum analyzer. 
\begin{figure}
  \includegraphics[scale=1]{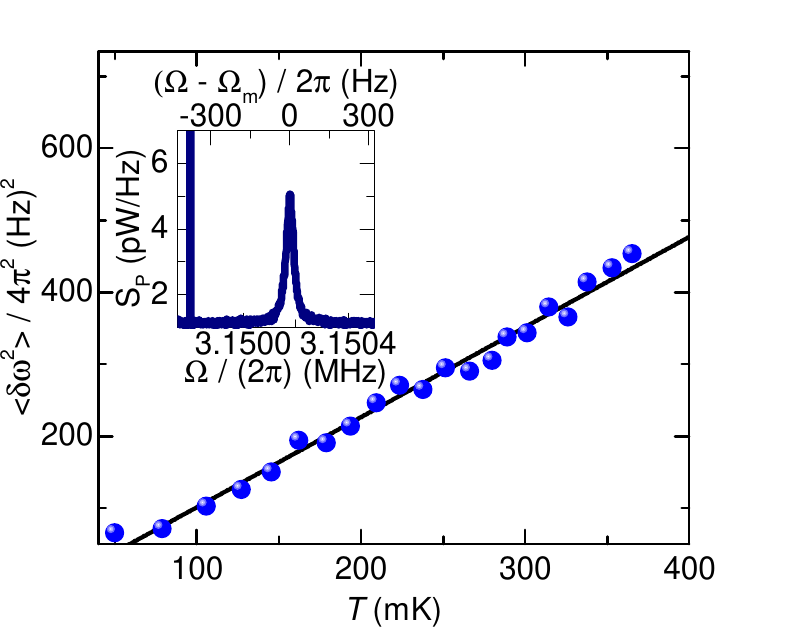}
        \caption{\textit{Thermal fluctuations of the nanostring plotted versus temperature.} The inset displays the noise power spectral density of the nanostring at $365\,$mK. The mechanical resonator has an eigenfrequency of $\Omegam / 2\pi = 3.1502\,$MHz and a quality of about $10^5$. The average fluctuations of the resonator are calculated by a reference peak (left). In the full frame these fluctuations are plotted versus the sample temperature (blue dots). The slope $s$, which yields the electromechanical vacuum coupling according to Eq.$\,$(\ref{eq:g0}), is determined from the linear dependency.}
        \label{fig:Beam}
\end{figure}
The inset of Fig.\,\ref{fig:Beam} shows the microwave resonator sideband noise spectroscopy data representing the thermal motion of the nanostring at $365\,$mK. We find a mechanical resonance frequency $\Omegam/2\pi= 3.15018$\,MHz with a linewidth of $\Gammam/2\pi = (33.5\pm0.1)$\,Hz, corresponding to a mechanical quality of about $Q_{\textrm{m}} \simeq 94~000$. The sharp peak on the left side originates from the frequency modulation of the probe tone (with a modulation frequency of $\Omega_{\textrm{mod}}/ 2\pi = 3.1498\,$MHz and a modulation depth of $\Omega_{\phi}/ 2\pi = 80\,$Hz) used for the calibration of the phase response of the microwave resonator (for details, see Refs.\,\onlinecite{Zhou2013, Hocke2012, Gorodetsky2010, Weber2016}). A quantitative comparison of this amplitude calibration peak $S_{\textrm{mod}}$ with the amplitude of the thermal motion peak $S_{\textrm{pp}}$ yields the integrated displacement noise $\langle\delta\omega^2\rangle$ and hence the vacuum coupling via \cite{Zhou2013, Hocke2012, Gorodetsky2010}:
\begin{equation} 
\langle\delta\omega^2\rangle = \int_{-\infty}^{\infty}S_{\omega\omega}(\omega)  \frac{d\omega}{2\pi} = \frac{\phi_0^2\Omega_{\textrm{m}}^2\Gamma_{\textrm{m}}}{4 \, \textrm{ENBW}}\cdot\frac{S_{\textrm{pp}}}{S_{\textrm{mod}}} = 2\gmV^2 \bar{n}_{\textrm{m}}.
\label{eq:w^2}
\end{equation}
Here, we use the measurement bandwidth ENBW $=1\,$Hz.
By repeating this measurement for various temperatures, we can calibrate the mechanical coupling with a finite back-action temperature ($\bar{n}_{\textrm{m}} \rightarrow T_{\textrm{ba}} + k_{\textrm{B}}T/ \hbar \Omegam$). Figure$\,$\ref{fig:Beam} shows the integrated displacement noise as a function of the temperature $T$. We find a slope of $s / (2\pi)^2 = (1.253\pm0.035)\,$kHz$^2/$K and hence a vacuum coupling of 
\begin{equation}
\gmV / 2\pi = \sqrt{\frac{s\hbar \Omegam}{k_{\textrm{B}}}} = (0.308\pm0.004) \, \textrm{Hz}.
\label{eq:g0}
\end{equation}

\section{Linewidth of the microwave resonator}
\begin{figure}
  \includegraphics[scale=1]{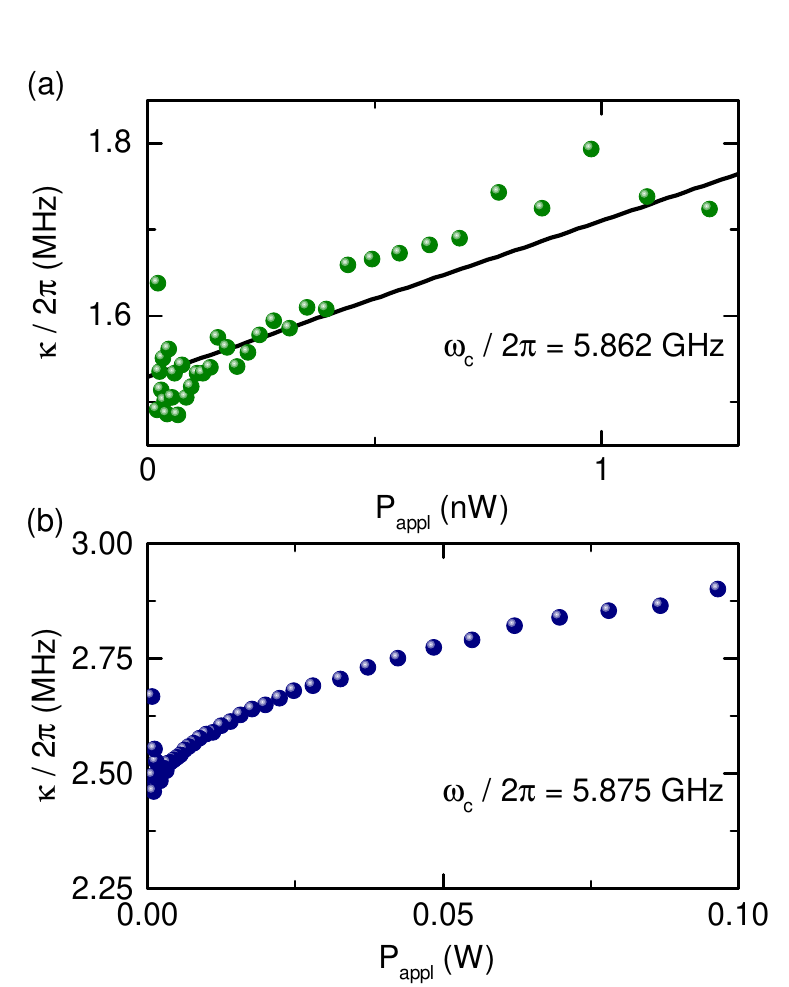}
        \caption{\textit{Dependence of the microwave resonator linewidth on the applied power.} Panel (a) displays the extracted linewidth in the qubit regime  (green dots) for a resonator frequency of $\omega / (2\pi) = 5.862\,$MHz [point A in Fig.\,\ref{fig:CavityResponse}(a)]. The applied power range is $0.02-1.2\,$nW. A linear trend is found, as indicated by the black solid line. Panel (b) shows the linewidth of the microwave resonator in the electromechanical regime (blue dots) for an applied power range of $0.9-97\,$mW applied power, where a non-trivial behavior is found. Here, the operation point of the transmon qubit is set to point B in Fig.\,\ref{fig:CavityResponse}(a).}
        \label{fig:ResonatorLinewidth}
\end{figure}

As the linewidth (loss rate) of the microwave resonator influences the average photon number in the resonator, see Eq.\,(1) in the manuscript, we analyze it for each of the two calibration methods. We note that for the two methods different working points of the transmon qubit are used, indicated by A and B in Fig.\,\ref{fig:CavityResponse}(a), resulting in different behaviors. The observed dependencies are depicted versus applied power in Fig.\,\ref{fig:ResonatorLinewidth} for the qubit (electromechanical) regime in panels (a) and (b), corresponding to the working points A and B, respectively.

The linewidth for the coupled qubit-resonator system is measured from $22\,$pW up to the critical photon number. We find a linear dependence with an offset of $(1.53\pm0.01)\,$MHz and a slope of $(181\pm14)\,$kHz/nW. For the calibration this linear dependence is used to interpolate the $\delta \omega \kappa^2$ product in Fig.\,3 of the manuscript.

A non-trivial behavior is observed in the electromechanical regime. After a peak at $0.9\,$mW, an increase up to $2.9\,$MHz is found at $\Papp = 97\,$mW. As the fluctuations in this dependence are rather small, we directly use the observed linewidth for our analysis. The measurement is done in-situ while determining the EMIA interference. We speculate that these fluctations of the linewidth arise from resistive elements in the Josephson junctions of the transmon qubit.

\bibliography{Bibliograph}
\end{document}